\documentclass{aa}

\usepackage{graphicx,natbib}
\usepackage{txfonts}
\bibpunct{(}{)}{;}{a}{}{,}

\begin{document}

\title{Non-Maxwellian electron distributions in clusters of galaxies}

\author{J.S. Kaastra\inst{1,2}
\and A.M. Bykov\inst{3}
\and N. Werner\inst{4}
}

\offprints{J.S. Kaastra}
\date{\today}

\institute{SRON Netherlands Institute for Space Research, Sorbonnelaan 2,
           3584 CA Utrecht, the Netherlands 
	 \and
	   Sterrenkundig Instituut, Universiteit Utrecht, 
           P.O. Box 80000, 3508 TA Utrecht, The Netherlands
	 \and
	   A.F. Ioffe Institute of Physics and
           Technology, St. Petersburg, 194021, Russia
	 \and
	   Kavli Institute for Particle Astrophysics and Cosmology,
	   Stanford University, 
	   452 Lomita Mall/mc 4085, Stanford, CA 94305, USA
	   }

\abstract
{ Thermal X-ray spectra of clusters of galaxies and other sources are commonly
calculated assuming Maxwellian electron distributions. There are situations
where this approximation is not valid, for instance near interfaces of hot and
cold gas and near shocks. }
{ The presence of non-thermal electrons affects the X-ray spectrum. To
study the role of these electrons in clusters and other environments, an
efficient algorithm to calculate the X-ray spectra is needed. }
{ We approximate an arbitrary electron distribution by the
sum of Maxwellian components. The decomposition is done using either a genetic
algorithm or an analytical approximation. The X-ray spectrum is then evaluated using 
a linear combination of those Maxwellian components. }
{ Our method is fast and leads to an accurate evaluation of the spectrum. The
use of Maxwellian components allows to use the standard collisional rates that
are available in plasma codes such as SPEX. We give an example of a
spectrum for the supra-thermal electron distribution behind a shock in a cluster
of galaxies. The relative intensities of the satellite lines in such a spectrum
are sensitive to the presence of the supra-thermal electrons. These lines can
only be investigated with high spectral resolution. We show that the instruments
on future missions like Astro-H and IXO will be able to demonstrate the presence
or absence of these supra-thermal electrons.}
{}


\titlerunning{Non-Maxwellian electron distributions}
\authorrunning{J.S. Kaastra et al.}

\maketitle

\section{Introduction}
\label{}

Most of the visible baryonic matter in clusters of galaxies is in the form of a
hot, tenuous gas. Usually it is assumed that this gas is in collisional
ionisation equilibrium. In the central parts the density is high enough and the
timescales long enough to fulfil the conditions for collisional ionisation
equilibrium. In the outer parts this may not always be the case. Freshly
accreted gas may still be ionising. For a review on equilibration processes in
such tenuous plasmas see \citet{bykov2008}. Nevertheless, usually still a
Maxwellian electron distribution is assumed to be valid. However, there are
situations where this is not an obvious constraint. For instance, in cooling
cores of clusters, electrons from hot gas may penetrate the colder parts. Also,
when shocks are present, deviations from a Maxwellian distribution may occur
associated with the temperature gradients or due to particles accelerated by the
shock. Shocks can be due to merger activity, AGN jets, and in cluster outskirts
accretion shocks may occur. As the Coulomb thermal relaxation times
increases with electron energy $E$ as $E^{3/2}$, supra-thermal electrons are
relatively long-lived and may yield a pressure contribution that is potentially
interesting for mass profiles. The presence of such supra-thermal electrons
can be revealed by excess emission in satellite lines.

In this paper we describe a way to model the emerging X-ray emission spectra for
the case of a non-Maxwellian electron distribution. Although our main focus is
here on clusters of galaxies, the same procedure can be applied to different
circumstances, for instance in stellar coronae, supernova remnants and the
Galactic ridge; for that last case see also \citet{masai2002} and
\citet{tatischeff2003}.

\section{X-ray spectra for supra-thermal electron distributions}

To calculate an X-ray spectrum, essentially two steps are needed: 1) determine
the ionisation balance and 2) calculate the corresponding spectrum. For more
details see e.g. \citet{kaastra2008}.

We first consider the ionisation balance. In order to determine that, it is
necessary to evaluate the collisional ionisation and recombination rates for
all ions. These rates are derived by integrating the relevant cross sections
over the electron distribution. For instance, the collisional ionisation rate
$C_{\mathrm{DI}}$ is given by
\begin{equation}
C_{\mathrm{DI}} = n_{\mathrm e}n_{\mathrm i}\int\limits_{0}^{\infty}
\sigma_i(E) \varv f(\varv){\mathrm d}\varv,
\label{eqn:collion}
\end{equation}
where $n_{\mathrm e}$ and $n_{\mathrm i}$ are the densities of electrons and the
ion $i$ considered, $\sigma_i(E)$ is the energy-dependent collisional ionisation
cross-section, $E=m_{\mathrm e}\varv^2/2$ is the kinetic energy of an electron
colliding with the ion and $f(\varv)$ is the electron velocity distribution.

In the most simple approximation, the cross section for collisional ionisation
can be written as \citep{lotz1967}:
\begin{equation} 
\sigma_i (E) = \frac{a n_{\mathrm s} \ln (E/I)}{EI}, 
\label{eqn:lotz}
\end{equation}
where $E$ is the kinetic energy of the free electron, $n_{\mathrm s}$ is the
number of electrons in a given atomic shell and the normalisation $a=4.5\times
10^{-24}$~m$^2$keV$^2$. The cross section (\ref{eqn:lotz}) can be inserted into
(\ref{eqn:collion}), and the integration can be done analytically for simple
electron distribution functions like Maxwellians or power laws, resulting in an
expression proportional to $E_1(I/kT)/IT^{0.5},$ where $I$ is the ionisation
potential of the relevant shell, and $E_1(x)=\int_x^{\infty}t^{-1}{\mathrm
e}^{-t}{\mathrm d}t$ is the exponential integral. Eq.~(\ref{eqn:lotz}) is
obviously too simple in realistic cases. But more sophisticated approximations
to the collisional ionisation cross section can be made, e.g.
\citet{younger1981}, that still allow analytical integration for Maxwellian or
power law electron distributions:
\begin{equation}
uI^2 \sigma_i (E) = A(1-1/u) + B(1-1/u)^2 + C \ln u + D \ln u / u,
\label{eqn:younger}
\end{equation}
with $u\equiv E/I$ and $A$, $B$, $C$ and $D$ adjustable parameters.
The use of analytical integration has the obvious advantage of enhanced
computational speed, as during the evaluation of a spectrum many atomic shells
of hundreds of ions need to be taken into account.

A similar treatment can be made for other relevant rates, like the  collisional
excitation rates that are important for the line emission. However, some rates
cannot be written in a form that allows analytical integration over a Maxwellian
electron distribution. A good example are the radiative recombination rates. We
recall that the cross section for radiative recombination $\sigma_{\mathrm{bf}}$
can be expressed in terms of the photoionisation cross section through Milne's
relation as
\begin{equation}
\sigma_{\mathrm{bf}}(\varv) 
= E_{\mathrm{ph}}^2 \sigma_{\mathrm{fb}}(E_{\mathrm{ph}}) / (m_{\mathrm e} c^2
m_{\mathrm e}\varv^2),
\label{eqn:milne}
\end{equation}
where $\sigma_{\mathrm{fb}}(E_{\mathrm{ph}})$ is the photoionisation cross
section at the photon energy $E_{\mathrm{ph}}=I+E$. Although there exist
analytical approximations to the photoionisation cross sections (for instance
\citealt{verner1995}), they are too complicated to allow for analytical
integration of (\ref{eqn:milne}) over a Maxwellian electron distribution, in
particular as they are more complicated than a simple power law near the
ionisation edges (Cooper minima). Even for the hydrogenic case, where the cross
section can be calculated analytically, its shape is too complex for analytical
convolution with the electron distribution. Therefore, the integration must be
done numerically. In practice, often the exact cross sections (including also
resonances near the edge) are convolved numerically with a Maxwell distribution,
and the resulting rates are then approximated by analytical functions.

It is clear that we need to follow an alternative approach for non-Maxwellian
electron distributions. There have been attempts to make a generalisation of the
Maxwell distribution. For instance, \citet{porquet2001} use a combination of a
Maxwellian at low energies with a power-law at high energies, but they only
consider the ionisation balance, not the resulting spectrum.
\citet{prokhorov2009} use a similar approximation for the electron distribution,
but only consider continuum emission and the \ion{Fe}{xxv} and \ion{Fe}{xxvi}
1s--2p blends. \citet{owocki1983} use the so-called kappa-distribution
\citep{olbert1967}:
\begin{equation}
f_\kappa(E) = 
\frac{ \Bigl( 
       \frac{ \displaystyle{m_{\mathrm e}} }
            { \displaystyle{2\pi kT} } 
       \Bigr)^{3/2}
       \frac{ \displaystyle{\Gamma(\kappa+1)} }
            { \displaystyle{(\kappa-3/2)^{3/2}\Gamma(\kappa-1/2)} } 
	       }{
       \Bigl( 
       1+\frac{ \displaystyle{E} }
              { \displaystyle{(\kappa-3/2)kT} }
       \Bigr)^{\kappa+1}
       }.
\label{eqn:kappa}
\end{equation}
This distribution tends to a Maxwellian for $\kappa\rightarrow\infty$ and it has
a power law tail proportional to $E^{-\kappa-1}$ for high energies. However, the
above approach has also clear disadvantages. Not every nonthermal electron
distribution can be cast into the shape of (\ref{eqn:kappa}), and not all rates
can be convolved analytically with (\ref{eqn:kappa}), in particular again the
recombination rates. \citet{owocki1983} had to assume very simple analytical
approximations to the cross sections, like (\ref{eqn:lotz}), in order to be able
to evaluate analytically the plasma rates, which are given in terms of
hypergeometric functions. Therefore we propose a different approach.

\section{The multi-Maxwellian approach}

Plasma codes generally may contain (ten)thousands of spectral lines, and at high
spectral resolution, spectra contain many bins. Therefore, numerical integration
for each rate over the electron distribution is a very time consuming task, and
not of practical use, in particular when spectral fitting is performed and the
model needs to be evaluated many times.

But all relevant rates in the case of the collisionally ionised plasmas that we
consider here, are proportional to the product $n_{\mathrm e}n_{\mathrm i}$, as
for each of these rates the interaction (collision) of an electron with an ion
is the driving process. Therefore, all rates are linearly proportional to the
electron density $n_{\mathrm e}$. If we decompose the electron distribution into
a linear combination of elementary components, the total rate for any process is
simply the sum of the rates for these individual elementary components. Now, for
Maxwellian electron distributions all rates are well known and fast to evaluate
analytically or using analytical approximations. Thus, if we decompose an
arbitrary electron distribution $f(E)$ into a linear combination of Maxwellians:
\begin{equation}
f(E) = \sum\limits_{i} A_i g_{\mathrm M}(E,kT_i),
\label{eqn:maxdec}
\end{equation} 
with $A_i$ the normalisation constant for component $i$ and $g_{\mathrm
M}(E,T_i)$ the Maxwell distribution for temperature $T_i$, then the
calculation of any rate (recombination, ionisation, excitation) is simple and
straightforward. The problem is then reduced to obtaining the normalisations
$A_i$ and temperatures $T_i$ for an arbitrary electron distribution $f(E)$. 

In principle, one might solve formally the integral equation
\begin{equation}
f(E) = \int\limits_0^{\infty}  A(kT) g_{\mathrm M}(E,kT)\, {\mathrm d}kT.
\label{eqn:maxint}
\end{equation}
Using \begin{equation}
g_{\mathrm M}(E,kT) = \frac{2E^{1/2}}{\pi^{1/2}(kT)^{3/2}}
e^{\displaystyle{-E/kT}}
\label{eqn:maxdis}
\end{equation}
(\ref{eqn:maxint}) can be cast into the shape of a Laplace transform as
\begin{equation}
\frac{\pi^{1/2} f(E)}{2E^{1/2}} = 
\int\limits_0^{\infty} 
B(y)
e^{-Ey}{\mathrm d}y,
\label{eqn:laplace}
\end{equation} 
with 
\begin{equation}
B(y) \equiv y^{-1/2} A(1/y).
\end{equation}
The calculation of the inverse Laplace transform is not always trivial. For
instance, when we consider a mono-energetic beam, $f(E)$ is essentially a
delta-function, and the formal solution is a sine-wave with infinite frequency,
which is not very practical. Fortunately, in most practical cases $f(E)$ is a
smooth function of energy, spanning a broad range of energies. 

There are various ways to solve (\ref{eqn:laplace}). There is software available
that can do the inverse Laplace transform numerically (for example
\citealt{damore1999}), but in practice this is difficult, because these
algorithms only work if the left-hand-side of (\ref{eqn:laplace}) is an
analytical function. However, in practice the electron distributions are given
in tabular form, and the algorithms used to interpolate such tables do not
produce a formal analytical function (the requirement that the function is
infinitely differentiable is usually violated). There are two solutions to this
problem that we elaborate in the next subsections: direct approximation of the
electron distribution by an analytical function, or direct decomposition of the
electron distribution as a sum of Maxwellians.

\subsection{Approximation of the electron distribution by an analytical
function\label{sect:analytical}}

In practice, apart from the most pathological cases, electron distributions
contain a core that is close to Maxwellian, plus a high energy tail. For an
example see Sect.~\ref{sect:example}. After some experimentation, we found
that the electron distribution of that example can be approximated by the
following series:
\begin{equation}
f(E) = \frac{x}{2kT_0} \Bigl[ c_0 e^{\displaystyle{-a_0 x^2}} 
+ \sum\limits_{k=1}^4
  \frac{c_k}{(a_k+x^2)^{2+k/2}} \Bigr],
  \label{eqn:polap}
\end{equation}
with $x$ the dimensionless momentum defined by
\begin{equation}
x \equiv \sqrt{E/kT_0},
\end{equation}
and where $T_0$ is a characteristic temperature corresponding to the electron
distribution, and the parameters $c_k$ and $a_k$ can be adjusted to give
the best fit.

Eq.~(\ref{eqn:polap}) is an analytical function, and the inverse Laplace
transform in this case can be done analytically, yielding
\begin{eqnarray}
A(kT) &=& \frac{c_0\sqrt{\pi}}{4a_0^{3/2}}\, \delta(kT-kT_0/a_0) \nonumber\\
&+& \sum\limits_{k=1}^{4} \frac{c_k}{kT_0 \Gamma(2+k/2)}\,\Bigl(
\frac{T_0}{T} \Bigr) ^{(3+k)/2}\,e^{\displaystyle{-a_kT_0/T}},
\label{eqn:lapsol}
\end{eqnarray}
where $\Gamma(x)$ is the $\Gamma$-function and $\delta(x)$ Dirac's
delta-function. 

For our example, the normalisations $c_0-c_4$ are  0.755, 0.0609, 2.54, 13.3,
and 17.58, respectively; the scale factors $a_0-a_4$ are  0.483, 152, 6843, 57.4
and 12.4. The fit is not perfect, but better than 0.5\% for $x<30$,  and better
than 1\% for $x<700$. The relative difference to the exact distribution are
shown in Fig.~\ref{fig:eldis}.

\subsection{Direct decomposition of the electron distributions 
as the sum of Maxwellians
\label{sect:genetic}}

We seek an approximation to $f(E)$ that can be written as a linear combination
of $n$ Maxwellians, with $n$ a given number. The solution can be represented as
a set of pairs $(T_i,A_i)$ with $T_i$ the temperatures and $A_i$ the
normalisations. We will choose for convenience a set of increasing temperatures
($T_{i+1}>T_i$), and we will demand that $A_i\ge 0$ (physically allowed
components). Most realistic electron distributions can be represented as a
Maxwellian with high-energy tails. Therefore, the first temperature $T_1$ should
be close to the temperature $T_0$ of this dominant Maxwellian component.

This then leads to the following approach. We define a set of $n$
logarithmically spaced temperature intervals $(T_{1,i},T_{2,i})$ with $T_{1,1}$
equal to $T_0$ and for all $i$, we take $T_{2,i}=sT_{1,i}$ and
$T_{1,i}=T_{2,i-1}$. For the parameter $s>1$ we adopt a value of 1.5. The total
temperature range spanned by these $n$ contiguous intervals is therefore $T_0 -
s^n T_0$. For each interval, we choose an arbitrary temperature $T_i$ within
this interval ($T_{1,i}\le T_i \le T_{2,i}$) and an arbitrary normalisation $A_i
\ge 0$. We scale the $A_i$ in such a way that their sum corresponds to the total
electron density. We then check how close the resulting electron distribution is
to the electron distribution $f(E)$ that we want to approximate. By varying all
values $T_i$ within their allowed ranges (the above mentioned intervals) and
also the values for $A_i$, we try to get the best possible approximation.

We use the genetic algorithm PIKAIA developed by \citet{charbonneau1995} to find
the best solution. This algorithm needs a formal function of the parameters
($T_i$, $A_i$) that needs to be maximised. For this function we choose the
inverse of the sum of the squared relative deviations of the approximation to
the true electron distribution $f(E)$.

After some experimenting, we obtained satisfactory convergence with the
following initial parameters: number of individuals 500, number of generations
500. All other parameters in PIKAIA were kept to their default values. The
number of individuals represents here the number of sets $(T_i,A_i)$ from which
the iteration starts; the number of generations is the number of iterations
during which the individual sets evolve towards the optimum solution.

We have tested our algorithm on various different electron distributions, but
give here only one illustrative example.

\section{Example of decomposition of an electron distribution into
Maxwellians\label{sect:example}}

\begin{figure}
\resizebox{\hsize}{!}{\includegraphics[angle=-90]{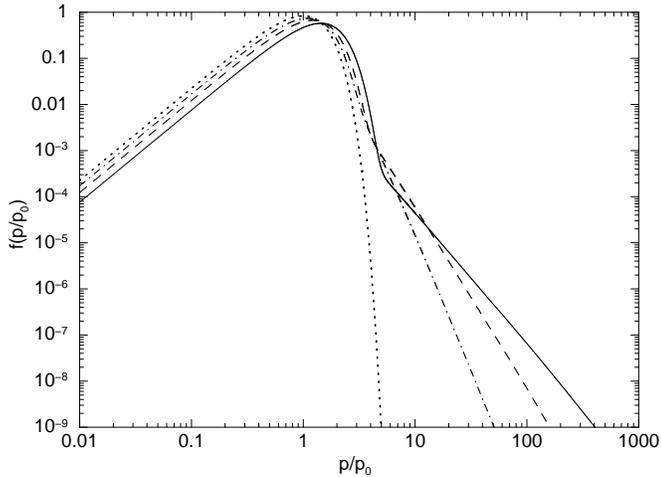}}
\caption{Electron distributions immediately after a shock with different Mach
numbers, in the downstream region.  The distributions are expressed here in
terms of the dimensionless momentum (normalised to the thermal momentum in the
far upstream region). The distributions are normalised to integral unity. Solid
line: $M=2.2$; dashed line: $M=1.5$; dash-dotted line: $M=1.2$; dotted line: far
upstream Maxwellian distribution.}
\label{fig:fp}
\end{figure}

\begin{figure}
\resizebox{\hsize}{!}{\includegraphics[angle=-90]{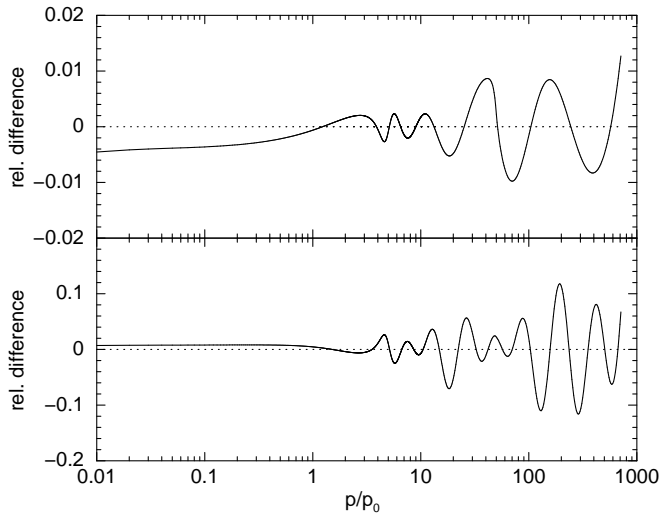}}
\caption{Relative differences of approximations to the electron distribution
with $M=2.2$ of Fig.~\protect\ref{fig:fp}.
Upper panel: analytical approximation
(\ref{eqn:polap}), see Sect.~\ref{sect:analytical}; lower panel: 
approximation using the sum of 32 Maxwellian components,
see Sect.~\ref{sect:genetic}.}
\label{fig:eldis}
\end{figure}

\begin{figure}
\resizebox{\hsize}{!}{\includegraphics[angle=-90]{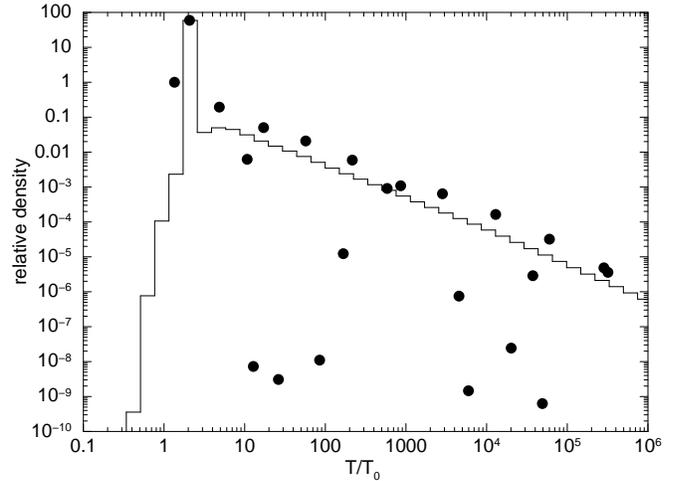}}
\caption{Relative normalisations versus temperature for the 32
Maxwellian components used in the approximation of the electron
distribution that is shown in Fig.~\ref{fig:eldis}. Histogram:
using the analytical approximation (\ref{eqn:lapsol}); Dots: using the 
solution solution from the genetic algorithm.}
\label{fig:decomp}
\end{figure}

We have used electron distributions based on the models of \citet{bykov1999} for
collisionless MHD shocks taking into account particle acceleration and a shock
precursor. We consider here a case for a pre-shock temperature of $T_0=10^8$~K,
electron density $10^{3}$~m$^{-3}$, magnetic field $10^{-10}$~T, and shock size
100~kpc. In the simulation we assumed a pure Maxwellian distribution of
electrons in the far upstream region of the flow and thus only the direct
injection of electrons from the upstream thermal pool to Fermi type shock
acceleration is responsible for the non-thermal electron population. The
electron distribution in the immediate post-shock region is shown in
Fig.~\ref{fig:fp} for three different Mach numbers. The model distribution was
calculated taking account of both Fermi-type acceleration in a collisionless
shock and Coulomb losses of the electrons in both the upstream and downstream
regions. It is normalised to the momentum $p_0$ corresponding to the typical
energy $kT_0$. Note that the supra-thermal electron distribution well away from
the shock front differs from that shown in Fig.~\ref{fig:fp}, because the
efficient Coulomb losses wash-out the non-relativistic supra-thermal electrons.

There is a variant of the model where mildly relativistic electrons (with
Lorentz factors $\gtrsim30$) comprising a putative long-lived cosmic ray
electron population in the ICM are re-accelerated by the MHD-shocks. The energy
efficiency problem is alleviated in that model in comparison with the case of
only direct particle injection from the thermal plasma. However, in this paper
we concentrate mostly on the diagnostic of non-relativistic electrons and thus
we consider only a direct injection model. 

For our example, we take the case of $M=2.2$. In our decomposition using the
genetic algorithm we have taken $n=32$ Maxwellians. The relative differences of
the approximation compared to the exact electron distribution is shown in
Fig.~\ref{fig:eldis}. The temperatures and relative normalisations of the
solution are shown in Fig.~\ref{fig:decomp}. We have also used the analytical
approximation (\ref{eqn:lapsol}) and binned it into similar temperature bins as
the solutions from the genetic algorithm.

\section{Spectrum for the supra-thermal electron distribution}

We have adjusted all models in the spectral fitting package SPEX that involve
emission or absorption from a hot plasma. All these models now include an option
to account for the presence of supra-thermal electrons. They have an additional
parameter, which is the name of a file containing the temperatures and relative
emission measures of the Maxwellian components. As SPEX does not use
pre-calculated tables but calculates spectra on the fly, all relevant rates
(ionisation, recombination, excitation) are simply calculated by adding the
contributions from the Maxwellian components. Obviously, this process is done in
two steps: first the composite multi-Maxwellian electron distribution is used to
determine the ionisation balance, and using the resulting non-equilibrium ion
concentrations, we calculate the X-ray spectrum for the non-equilibrium electron
distribution. It is tacitly assumed here that we consider a time-independent,
steady situation.  For the example given in the previous section, only a few
dozen Maxwellian components are needed. This allows fast and accurate evaluation
of the spectrum, without the need to make simplifications to the atomic
physics. 

The most important effects of non-thermal electrons on the spectrum is then a
shift of the ionisation balance towards higher ionisation, the production of a
non-thermal Bremsstrahlung tail on the continuum spectrum, and enhanced
satellite line emission. These satellite lines (for instance in the Fe-K band)
can be detected easily with high-resolution spectrometers that will fly on
future missions such as Astro-H and IXO. Their relevance as indicators for the
presence of non-thermal electrons was already indicated by \citet{gabriel1979}.

\begin{figure}
\resizebox{\hsize}{!}{\includegraphics[angle=-90]{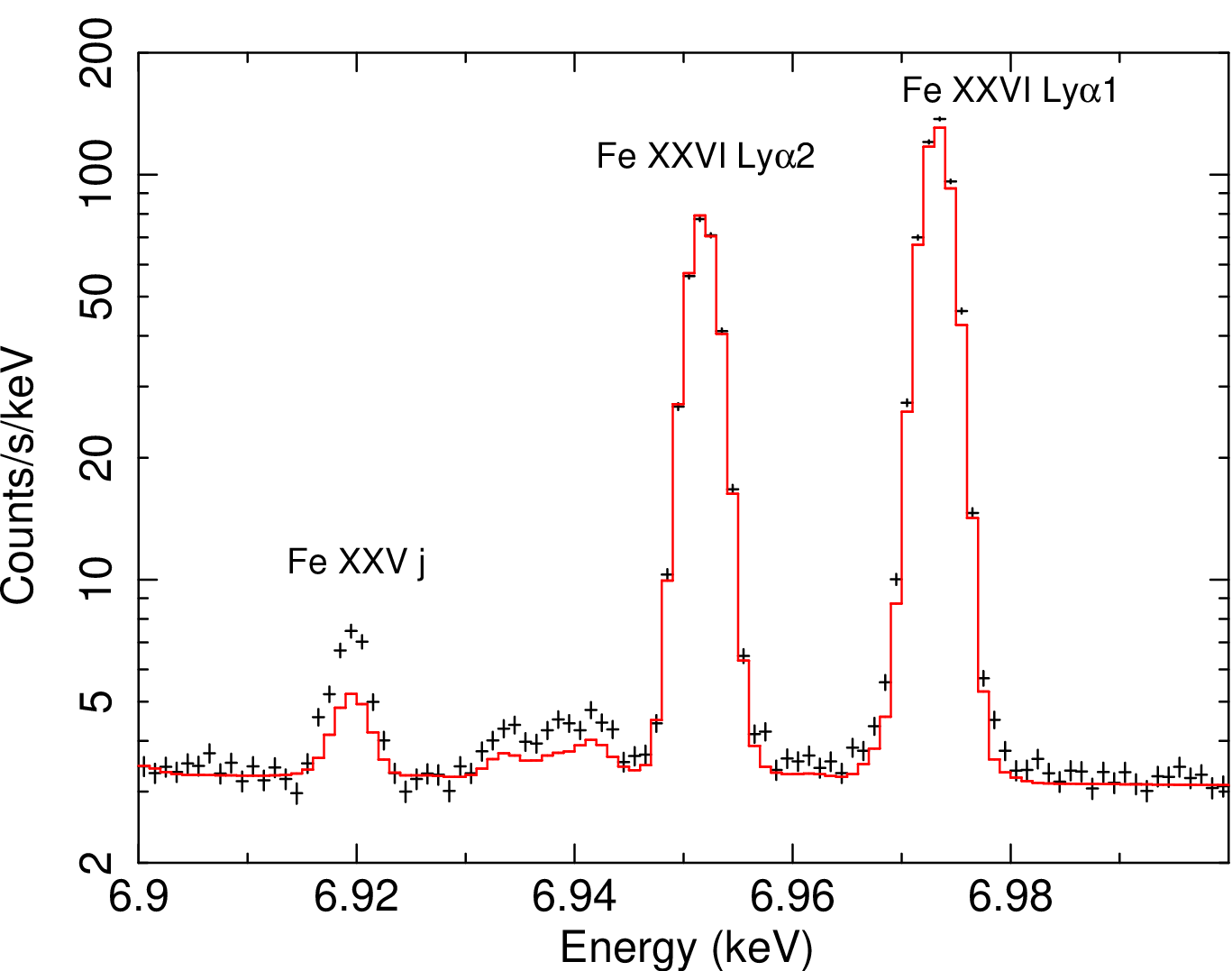}}
\resizebox{\hsize}{!}{\includegraphics[angle=-90]{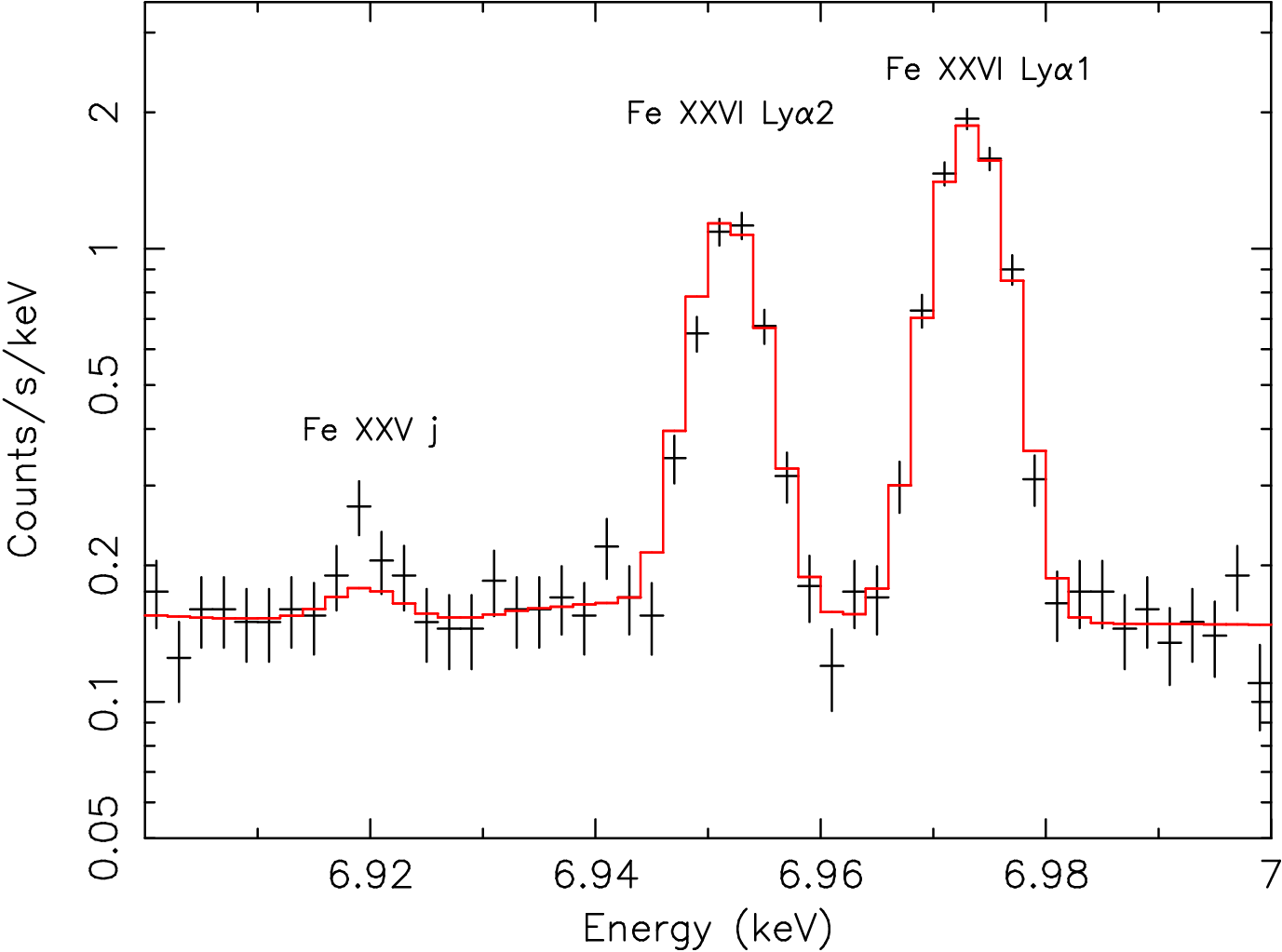}}
\caption{Crosses: simulated 100~ks calorimeter spectrum for IXO (top panel) and
Astro-H (bottom panel) as described in the text, for the supra-thermal electron
distribution of Fig.~\ref{fig:fp}. Solid line: best-fit model to a pure
Maxwellian plasma, with temperature 16.99~keV. Note the excess emission of
satellite lines in the data, in particular the \ion{Fe}{xxv} $j$-line.}
\label{fig:ixo}
\end{figure}

To illustrate the effects of such a supra-thermal electron distribution on data,
we simulated an XMM-Newton EPIC/pn spectrum extracted from a circular region
with a radius of 1\arcmin\ centred on the core of a bright cluster with a
$0.3-10$~keV luminosity of $L_{\mathrm{X}}=6.3\times10^{37}$~W within the
extraction region, at an assumed redshift of $z=0.055$. In the simulation of the
spectrum, we assumed the above mentioned post-shock downstream electron
distribution for the Mach number of $M=2.2$ and pre-shock temperature of
$kT=8.62$~keV ($10^8$~K). We assume a deep 100~ks observation. The resulting
spectrum has very high statistics, which should in principle allow to detect any
non-isothermality of the plasma. The best fit temperature of the simulated
spectrum is $17.86\pm0.10$~keV ($2\times10^8$~K), consistent with the post-shock
temperature of the plasma given by the Rankine-Hugoniot jump condition. An
isothermal model fits the data extremely well (reduced $\chi^2=1.02$) and the
non-thermal tail of the electron distribution cannot be detected in the
spectrum. 

We also simulated a spectrum with the same input parameters as observed during a
deep 100~ks observation with the X-ray micro-calorimeter on the proposed
International X-ray Observatory (IXO). We fitted the simulated spectrum with an
isothermal model and obtained a temperature of $kT=16.99\pm0.03$~keV, about
1~keV lower than the expected post-shock temperature. In Fig.~\ref{fig:ixo} we
show the $6.9-7.0$~keV part of the spectrum (rest-frame energies) which shows
the \ion{Fe}{xxvi} Ly$\alpha$ lines and the \ion{Fe}{xxv} $j$-satellite line.
Enhanced equivalent widths of satellite lines are good indicators of non-thermal
electrons. The satellite line in the simulated spectrum in Fig.~\ref{fig:ixo} is
clearly stronger than that predicted by the thermal model with a Maxwellian
electron distribution. This exercise illustrates, that in order to
observationally reveal non-Maxwellian tails in the electron distributions, we
will need high-resolution spectra obtained by future satellites with a large
effective area. 

However, even before IXO, Astro-H (expected launch 2014) will be able to detect
supra-thermal electrons. We simulated the same spectrum for the same extraction
region as above for the main instruments of Astro-H, again for 100~ks exposure
time. The two hard X-ray telescopes detect the source up to $\sim 75$~keV, and
the spectrum is well approximated ($\chi^2=225$ for 223 degrees of freedom) by
an isothermal model with measured temperature of $17.64\pm 0.12$~keV. Thus, the
presence of such a small amount of supra-thermal electrons cannot be revealed as
a hard tail, but it can be revealed in high-resolution spectra.
Fig.~\ref{fig:ixo} shows the simulated spectrum for the Soft X-ray Spectrometer
(SXS) of Astro-H. The excess flux at the \ion{Fe}{xxv} $j$-satellite has a
$3\sigma$ significance. Obviously, longer exposure times will enhance the
significance.

Interestingly, in all our simulations above, the best-fit iron abundance
for the isothermal model is about 30~\% higher than the actual abundance
that we have put into our model spectrum with supra-thermal electrons. This
holds also if we restrict our fit only to the Fe L-shell or Fe K-shell band.
This is because the higher-energy electrons have less efficient line
emissivity relative to the continuum, compared to lower-energy electrons.
Without knowing the amount of supra-thermal electrons, which can be
determined only from high-resolution spectra, this abundance bias cannot be
resolved and will result into incorrect interpretations.

\section{Concluding remarks}

In practice most effort goes into finding a good decomposition of an electron
distribution into Maxwellians. We have indicated and illustrated in this paper
two different methods: fits to simple analytical models that allow analytical
inversion of the Laplace transform, and a direct decomposition using a genetic
algorithm.

Finally we note that all relevant plasma rates that are used in the SPEX code 
are calculated using non-relativistic approximations. For instance, ionisation
cross sections are approximated by analytical functions like (\ref{eqn:younger})
that loose their validity for relativistic energies. Thus, for electron
distributions containing a significant fraction of relativistic electrons,
the results will be less accurate. Fortunately, in most situations this is
not a problem. For instance, the electron distribution of Fig.~\ref{fig:fp}
has a high-energy tail roughly proportional to $p^{-3}$. Inserting this for
instance into (\ref{eqn:collion}) and using (\ref{eqn:lotz}) for the high-energy
limit, shows that the integrand scales, apart from a logarithmic term,
proportional to $E^{-3}$, and therefore the highest energy electrons do not
contribute much to the rates. Therefore, even though the approximations made
to the decomposition of the electron distribution, in particular the genetic
algorithm, are not always very accurate at high energies (see
Fig.~\ref{fig:eldis}), this affects the final spectrum to a much lesser extent.

Only in astrophysical situations with a significant number of relativistic
electrons our method will not apply. \citet{bykov2002} has given an example of
this in the context of supernova remnants, considering only line fluorescence
due to collisional ionisation. Good approximations to the relativistic
collisional ionisation cross section of K-shell and L-shell electrons are
available (see references in \citealt{bykov2002}), but for a full plasma model
relativistic corrections to all rates would be needed, which is beyond the scope
of the present paper.

\begin{acknowledgements}

The Netherlands Institute for Space Research is supported financially by NWO,
the Netherlands Organization for Scientific Research. A.M.~Bykov was supported
by RBRF grant 09-02-12080. Support for this work was provided by the National
Aeronautics and Space Administration through Chandra Postdoctoral Fellowship
Award Number PF8-90056 issued by the Chandra X-ray Observatory Center, which is
operated by the Smithsonian Astrophysical Observatory for and on behalf of the
National Aeronautics and Space Administration under contract NAS8-03060.

\end{acknowledgements} 
 
\bibliographystyle{aa}
\bibliography{nonmax}

\end{document}